\documentclass{iau}

\usepackage{amsmath}
\usepackage{graphicx}
\usepackage{multirow}
\usepackage{aas_macros}
\usepackage{lipsum}

\usepackage{yfonts}
\usepackage{xcolor}

\begin{document}

\lefttitle{K. Vida et al.}
\righttitle{Searching for stellar CMEs in the Praesepe and Pleiades clusters}

\jnlPage{1}{7}
\jnlDoiYr{2021}
\doival{10.1017/xxxxx}

\aopheadtitle{Proceedings IAU Symposium}
\editors{
Nat Gopalswamy, Olga Malandraki, Aline Vidotto \& Ward Manchester eds.}

\title{Searching for stellar CMEs in the Praesepe and Pleiades clusters}

\author{K. {Vida}$^{1,2}$, B. {Seli}$^{1,2,4}$,  R. M. {Roettenbacher}$^3$, A. {G\"orgei}$^{1,2,4}$,  L. {Kriskovics}$^{1,2}$,  Zs.~{K{\H{o}}v{\'a}ri}$^{1,2}$, K. {Ol\'ah}$^{1,2}$}
\affiliation{
 \email{vidakris@konkoly.hu}\\
 $^1$ Konkoly Thege Mikl\'os Astronomical Institute, HUN-REN Research Centre for Astronomy and Earth Sciences\\
$^2$ CSFK, MTA Centre of Excellence \\
$^3$ University of Michigan\\
$^4$ E\"otv\"os Lor\'and University}

\begin{abstract}
On the Sun, the energetic, erupting phenomena of flares and coronal mass ejections (CMEs) often occur together. While space-based photometry has revealed frequent white-light flares for vast numbers of stars, only a handful of coronal mass ejections have been detected. Space-based photometry reveals the timing and detailed structure of flares. To detect CME signatures, however, optical spectroscopy is essential, as the ejected plasma can be detected by Doppler-shifted emission bumps in the Balmer-regions. We present a dedicated ground-based multi-object spectroscopic observations of the young, nearby Praesepe (600 Myr) and Pleiades (135 Myr) clusters to detect CMEs and flares parallel with the observations of Praesepe by the TESS satellite. During the 10 days of overlapping observations, we did not find any obvious signs of CMEs or flares in the H$\alpha$ region.
\end{abstract}


\maketitle

\section{Introduction}

%

Flares and coronal mass ejections (CMEs) are powerful events that occur on the surface of stars, including the Sun. Flares are sudden, explosive releases of energy resulting from the twisting and reconnection of surface magnetic fields. This process releases bursts of radiation across various wavelengths, from radio waves to X-rays. On the Sun, most large flares are accompanied by massive eruptions of plasma known as CMEs. Assuming that solar and stellar activity signatures are driven by the same underlying physics, a similar relationship between flares and CMEs is expected on other stars. These events significantly impact the space environment around the star by producing energetic particles, accelerating charged particles to high speeds, and disrupting the interplanetary magnetic field. 
Large flares and accompanying CMEs may render planets uninhabitable  
\citep{Vida2017ApJ...841..124V, 2017ApJ...851...77R}
 or lead  to the total loss of the planetary atmosphere  
 \citep{2007AsBio...7..167K, 2011OLEB...41..503L}. Understanding these phenomena is important for studying the behavior of stars, as well as for predicting and mitigating their effects on space weather and potential habitability. 

On the Sun,  flares and CMEs are studied by dedicated monitoring, partly motivated by the immediate threat they pose to the Earth. On other stars, however, flares are much better characterized than CMEs, since the latter are challenging to observe. Most of the known stellar CMEs were detected serendipitously using optical spectroscopy when the Doppler-shifted signal of the ejecta caused extra emission 
in the blue wing of Balmer lines  
(see, e.g., \citealt{1990A&A...238..249H} ). Currently, almost all known CME events are detected on M-dwarfs  
(see \citealt{2019A&A...623A..49V, Koller2021A&A...646A..34K, Lu2022A&A...663A.140L} and references therein).
Even though there are numerous detections of superflares on solar-like stars 
\citep{Maehara2012Natur.485..478M},
the only one exhibiting signatures of a CME is the young solar analog EK Dra, where  simultaneous spectroscopic and TESS photometry revealed evidence of a CME associated with a superflare 
\citep{2021NatAs...6..241N}.
Despite these advances, no comprehensive study of CME parameters and rates can be carried out. This poses a problem since CMEs have a large impact on the angular momentum evolution of young stars and can affect the atmospheres of exoplanets. Furthermore, current detections are based on either single-target observations, serendipitous detections, or archival data with a sub-optimal observation strategy for this purpose. 
A good solution could be observing open clusters, where several stars can be studied simultaneously -- such multi-object observations have been performed by \citet{Guenther1997, Leitzinger2014, Korhonen2017}.
In this manuscript, we present an attempt to observe stellar CMEs in a 
sample of young stars in open clusters using contemporaneous ground-based spectroscopic and space-based photometric observations.

\section{Target selection}
\begin{figure}
\centering
    \includegraphics[width=.45\textwidth]{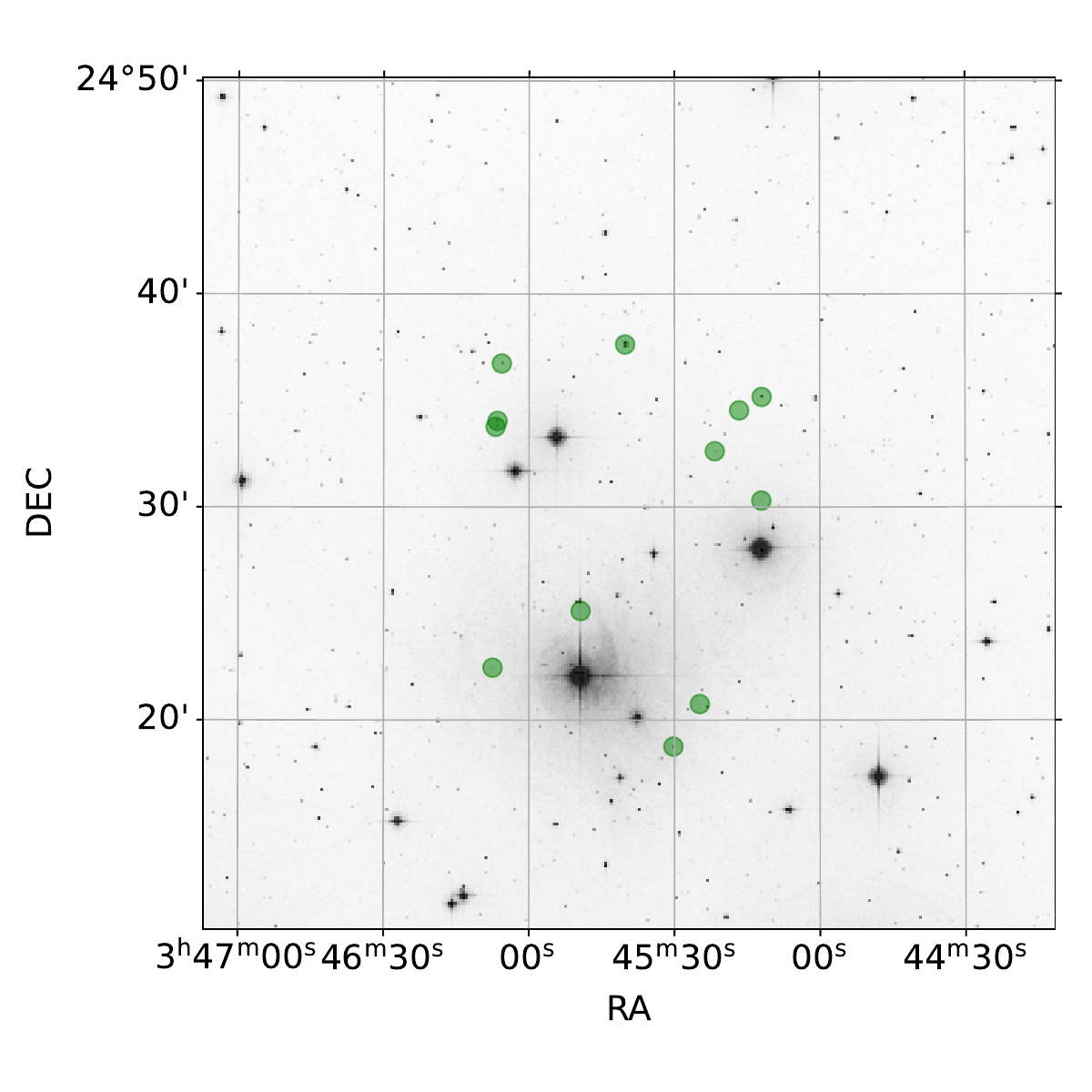}%
        \includegraphics[width=.45\textwidth]{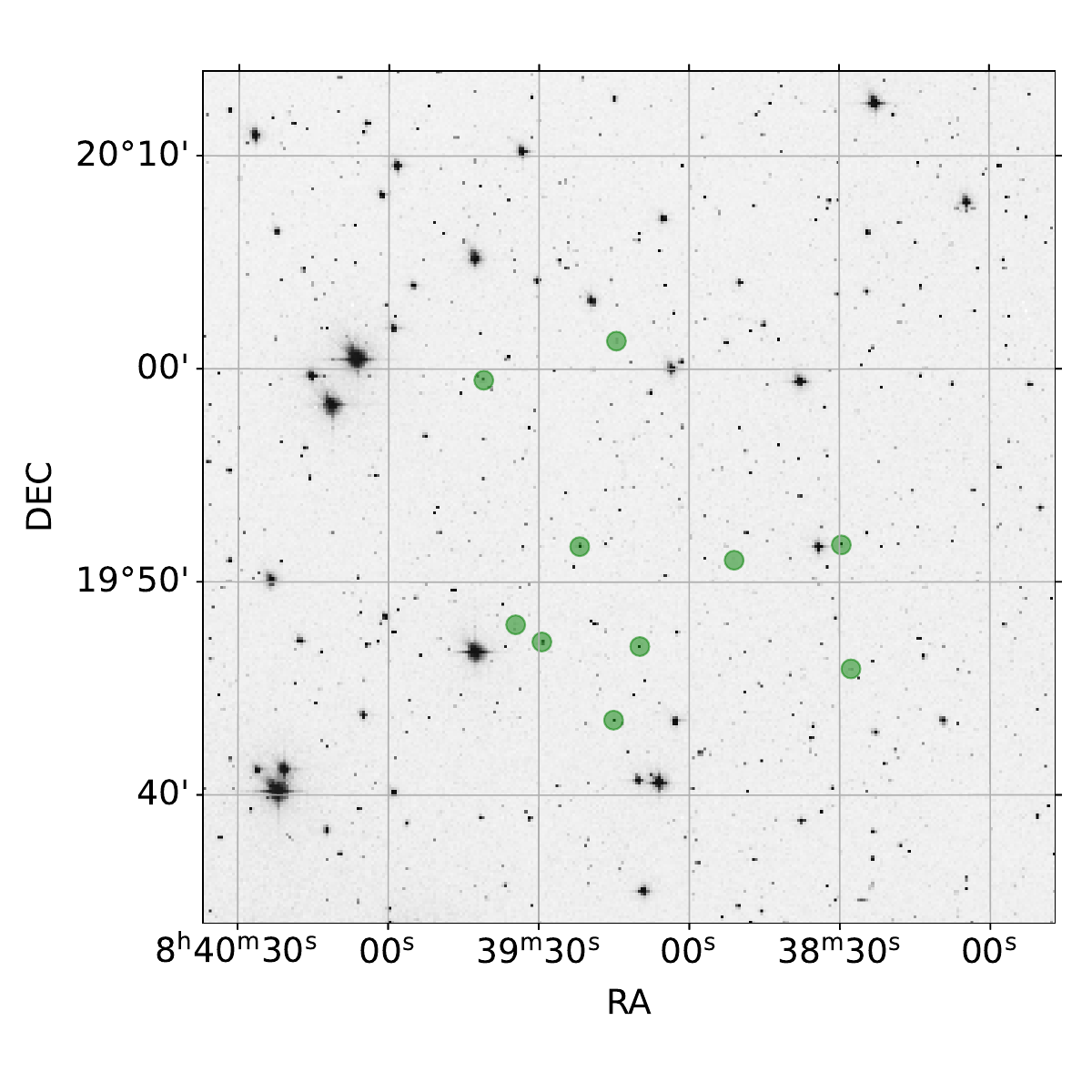}
    \caption{The field of view in the Pleiades (left) and Praesepe clusters (right) with the selected spectroscopic targets marked.}
    \label{fig:FoV}
\end{figure}

We selected two young, nearby open clusters: Pleiades (M45) and Praesepe (M44),  with ages of 135 and 600 Myr 
\citep{2018ApJ...863...67G}, respectively. 
To maximize our chances of detecting a CME, we need to observe as many simultaneous targets as possible. Therefore, our best option is to pick a young open cluster with numerous magnetically active stars that are most likely to exhibit (super)flares and CMEs. To achieve a high S/N ratio, we need to select a nearby cluster.  There are only three clusters that are suitable for our purposes: Hyades, Praesepe, and Pleiades. From these candidates, Praesepe and Pleiades were the best candidates for our purposes, as Hyades is the least dense cluster, and we could fit only half as many stars in a single FoV.

We analyzed photometric observations of the cluster members obtained earlier by the K2 mission and TESS to determine the chances of flaring in the possible fields. 
We selected targets to observe from the catalog of 
\cite{2021A&A...645A..42I}, who searched for stellar flares in open clusters with Kepler K2 photometry. We selected flaring stars brighter than $T=15^m$ (TESS magnitude) from Praesepe and Pleiades. We carried out a grid search in RA--Dec to find the optimal FoV with an 18' diameter, which contains the most flaring stars. With the best setup, we could include 10 stars from Praesepe and 12 from Pleiades (see Figure \ref{fig:FoV}).

\section{Early results}
\begin{figure}
\centering
\includegraphics[width=.30\textwidth]{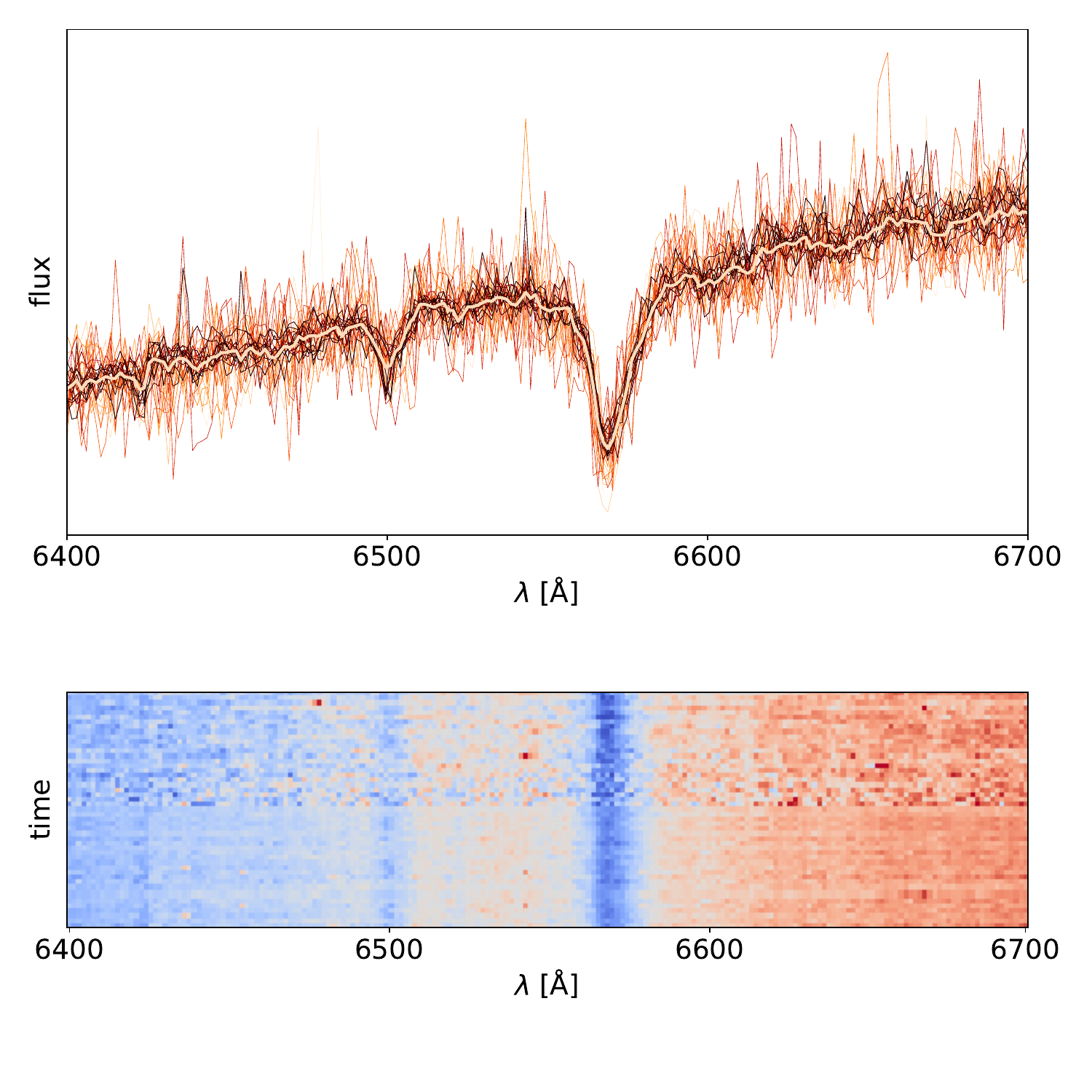}%
\includegraphics[width=.30\textwidth]{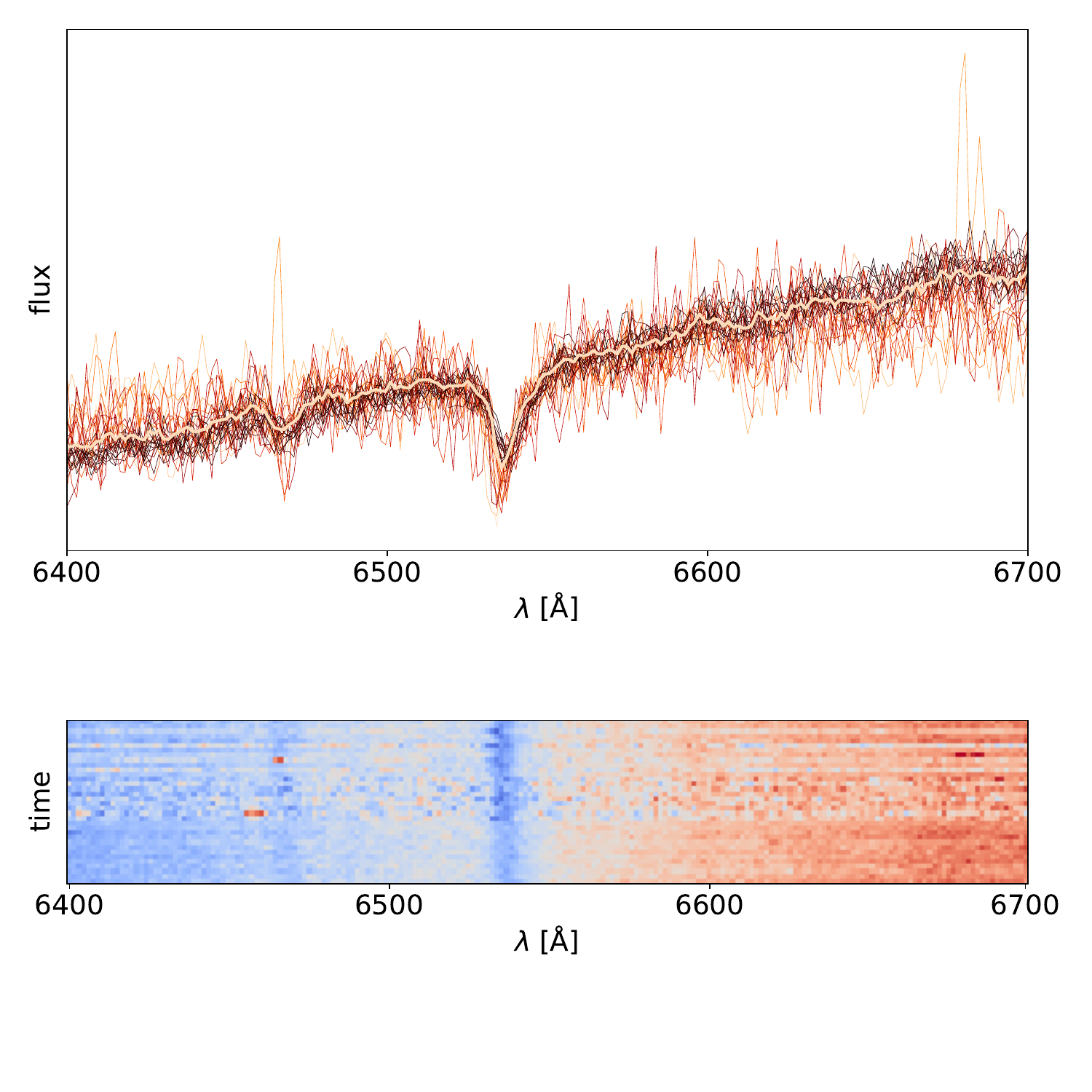}
\includegraphics[width=.30\textwidth]{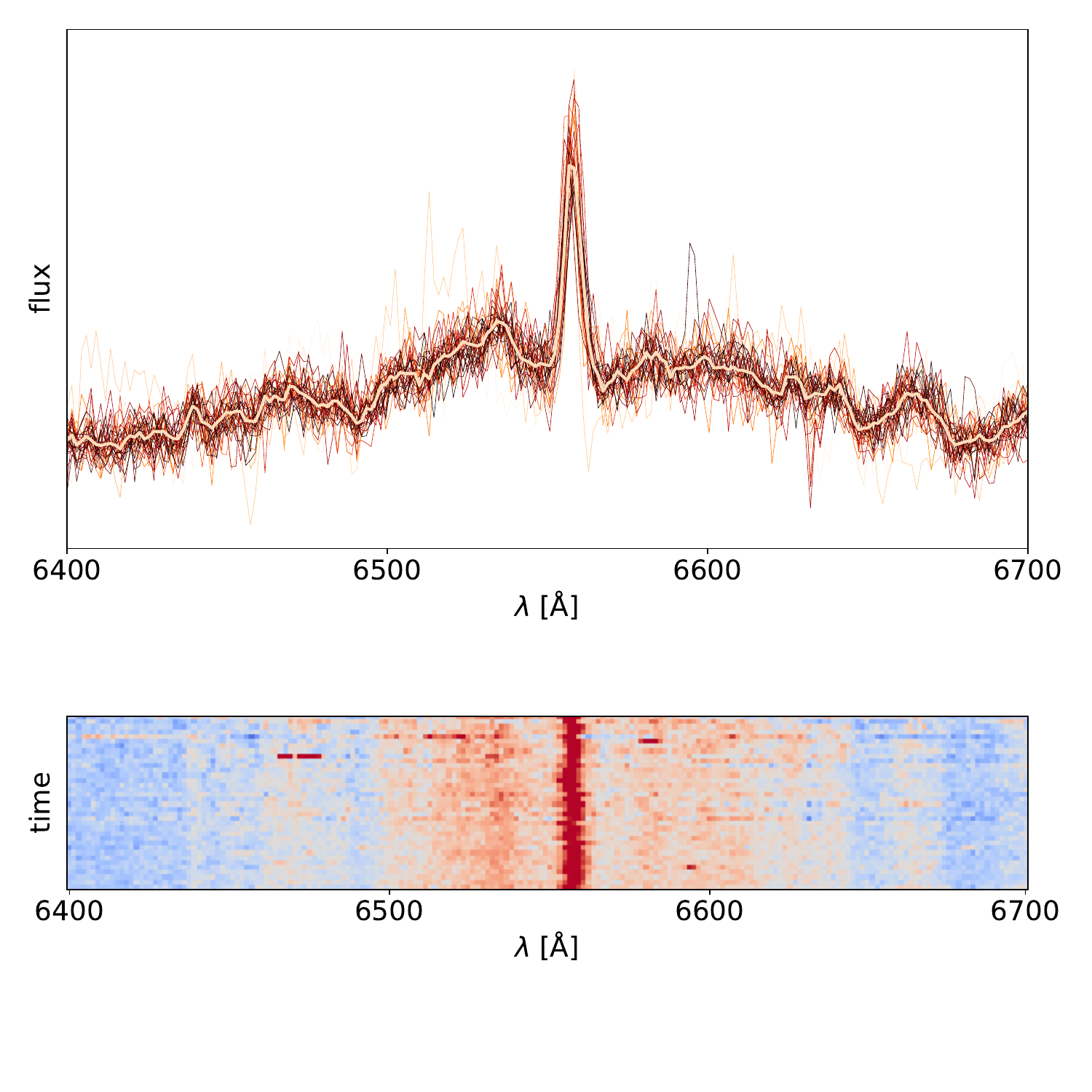}%
\includegraphics[width=.30\textwidth]{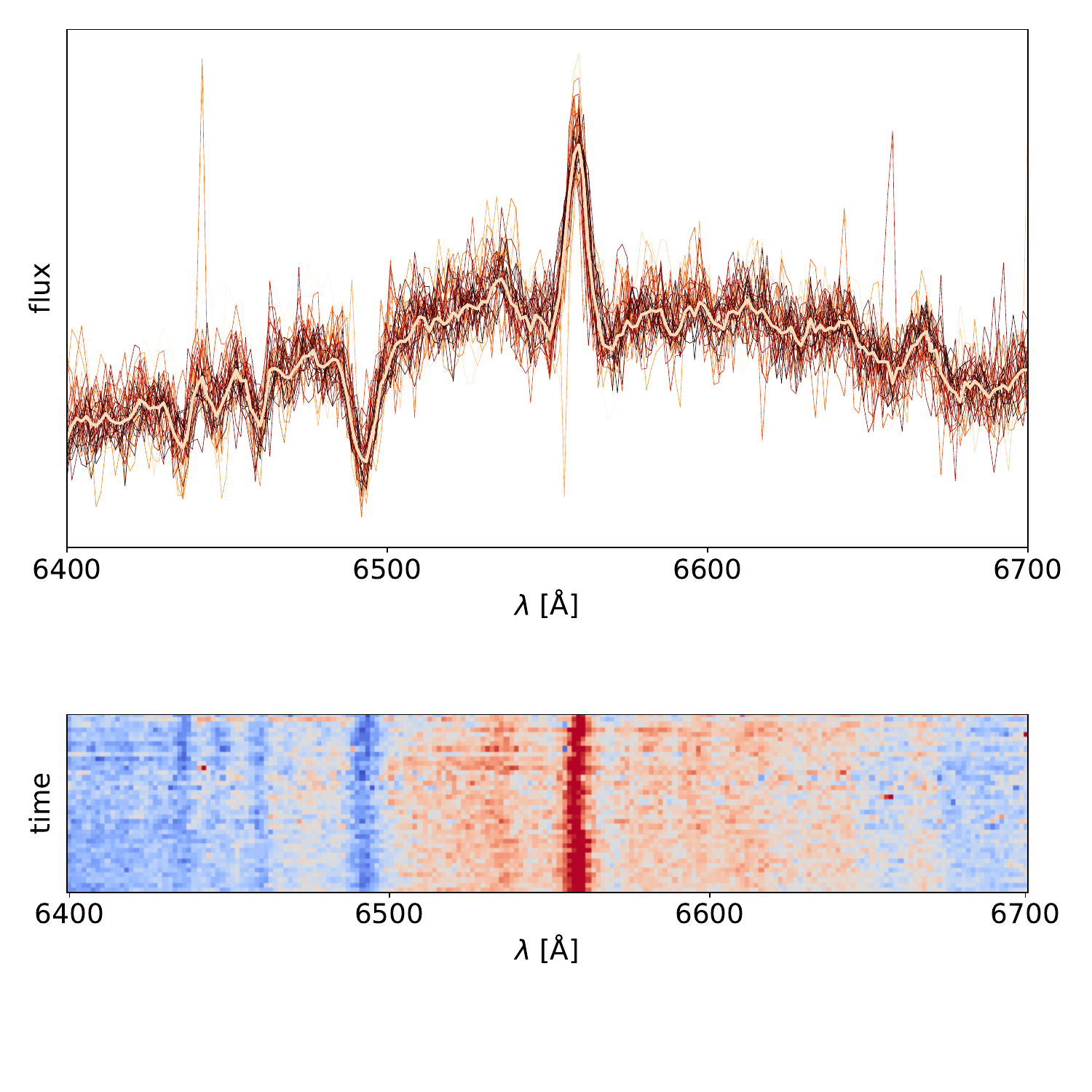}

    \caption{Example dynamic spectra and dynamic spectra from the Pleiades cluster showing the H$\alpha$ region. The continuous thick line indicates the averaged spectra.}
    \label{fig:sample-pleiades}
\end{figure}

\begin{figure}
\centering
\includegraphics[width=.30\textwidth]{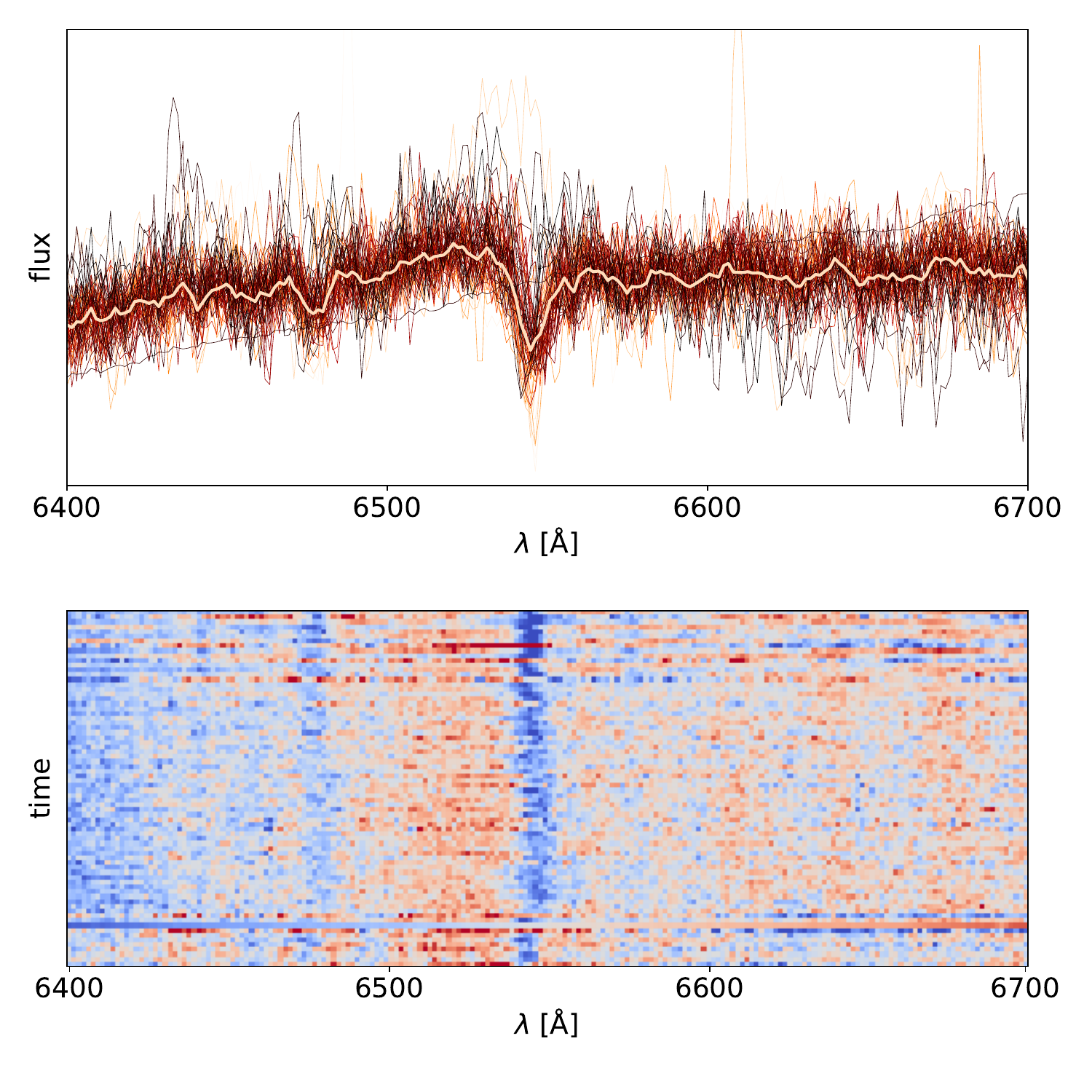}%
\includegraphics[width=.30\textwidth]{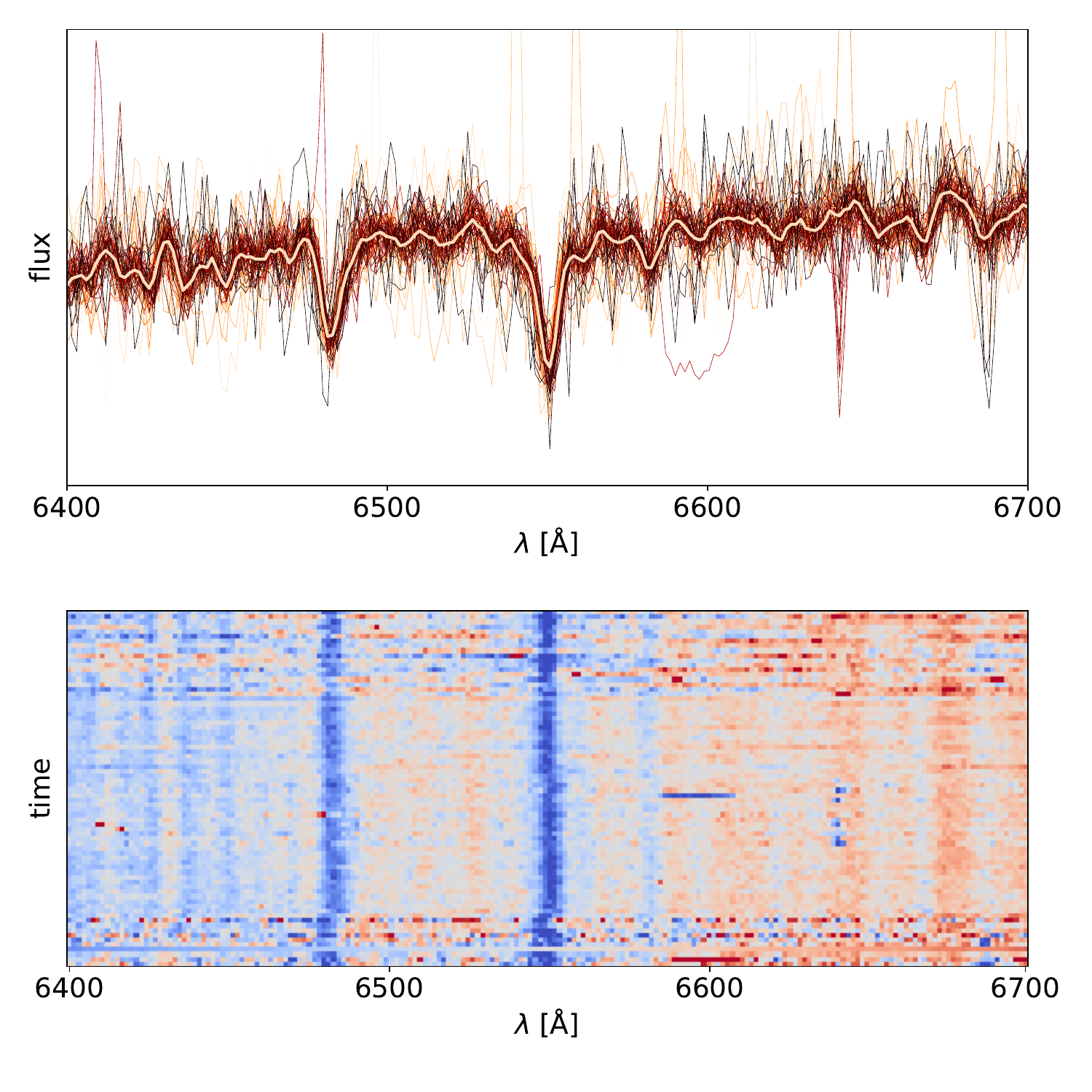}%
\includegraphics[width=.30\textwidth]{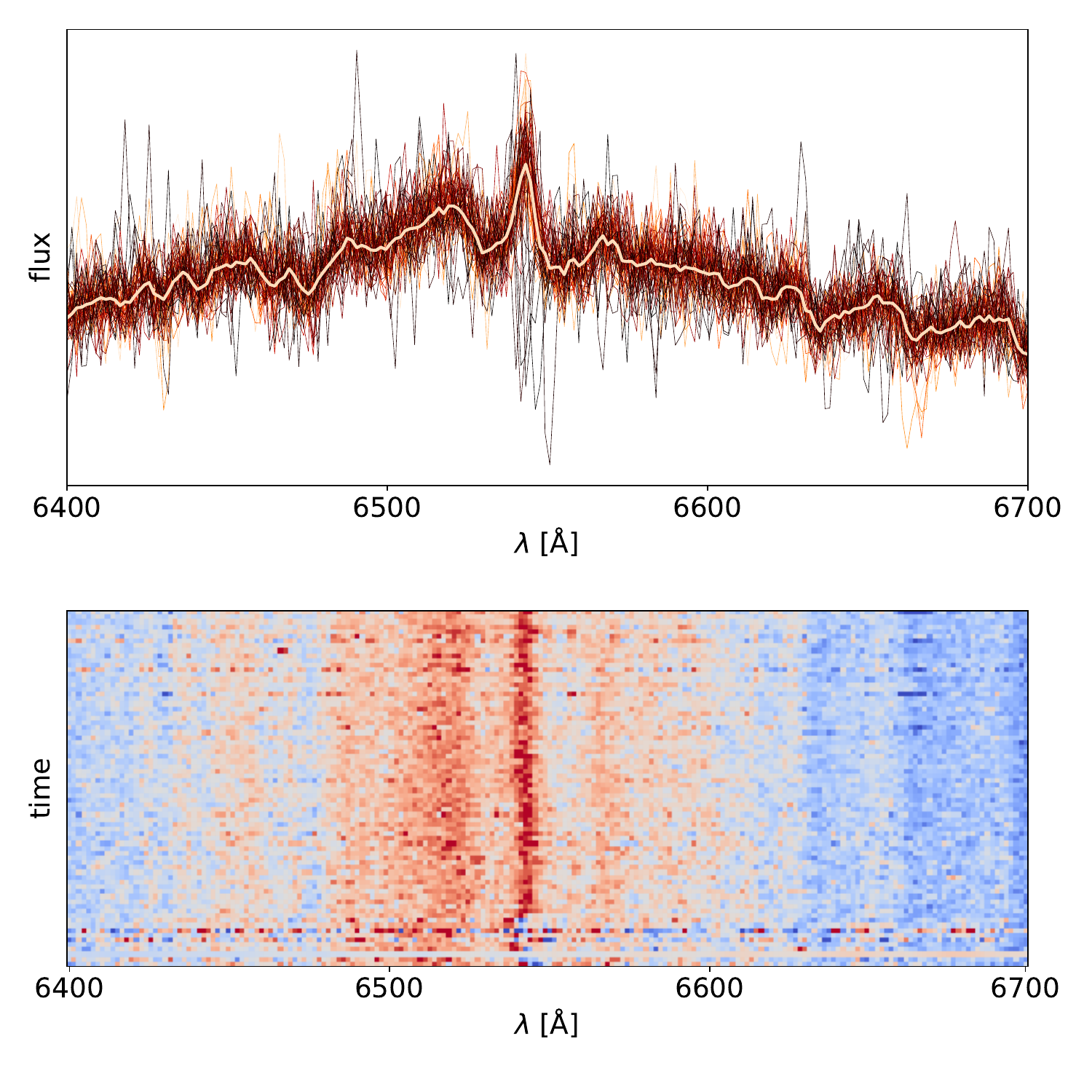}
\includegraphics[width=.30\textwidth]{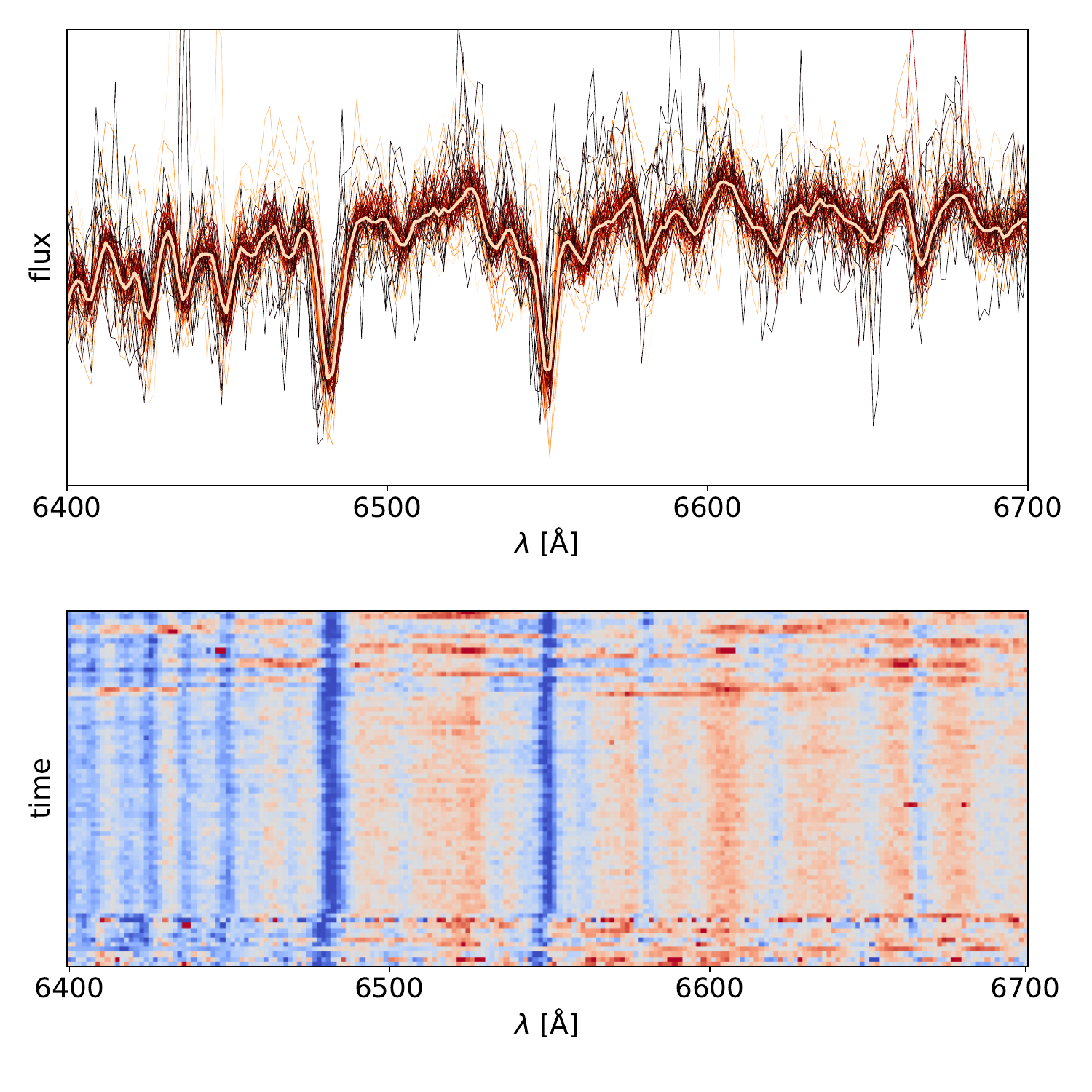}%
\includegraphics[width=.30\textwidth]{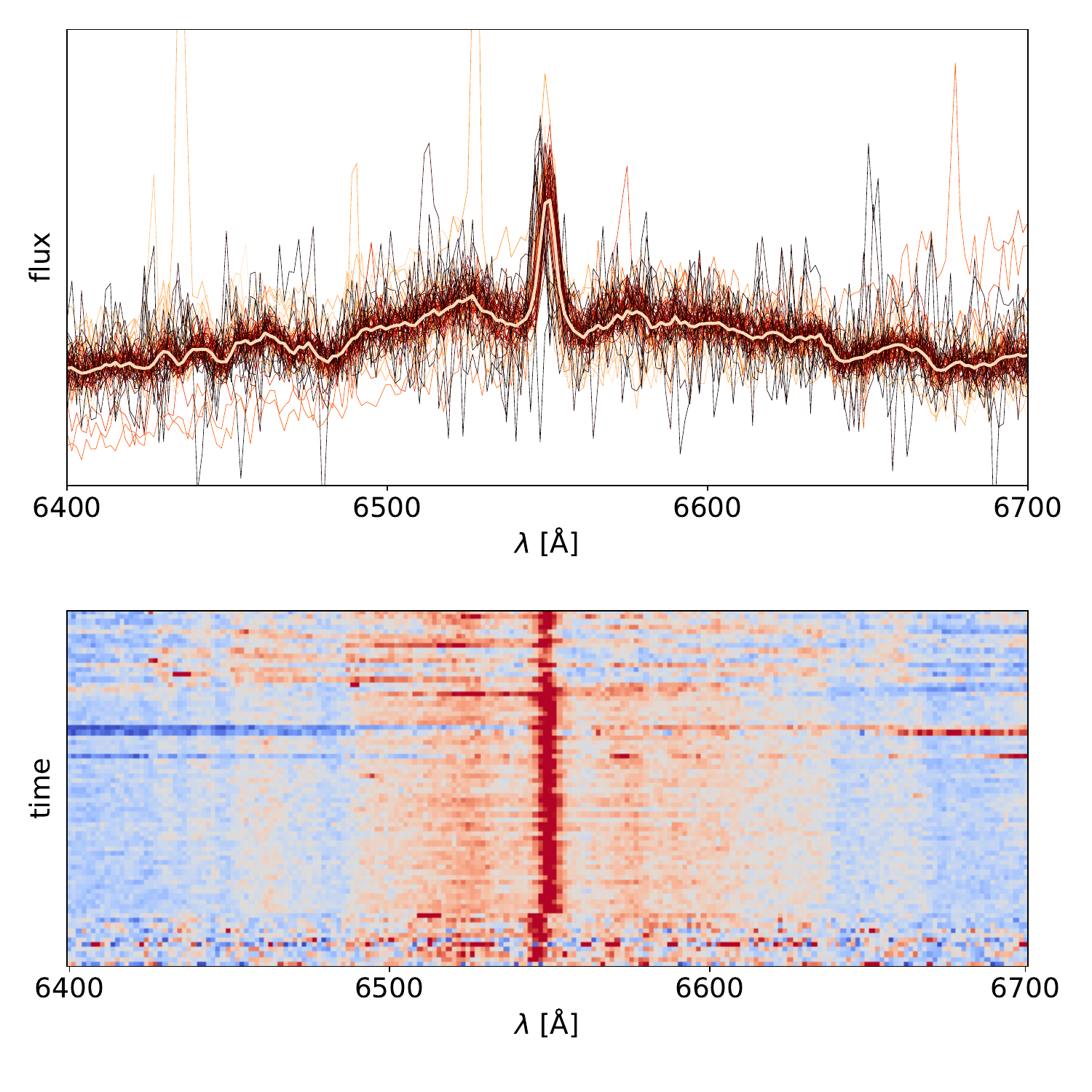}%
\includegraphics[width=.30\textwidth]{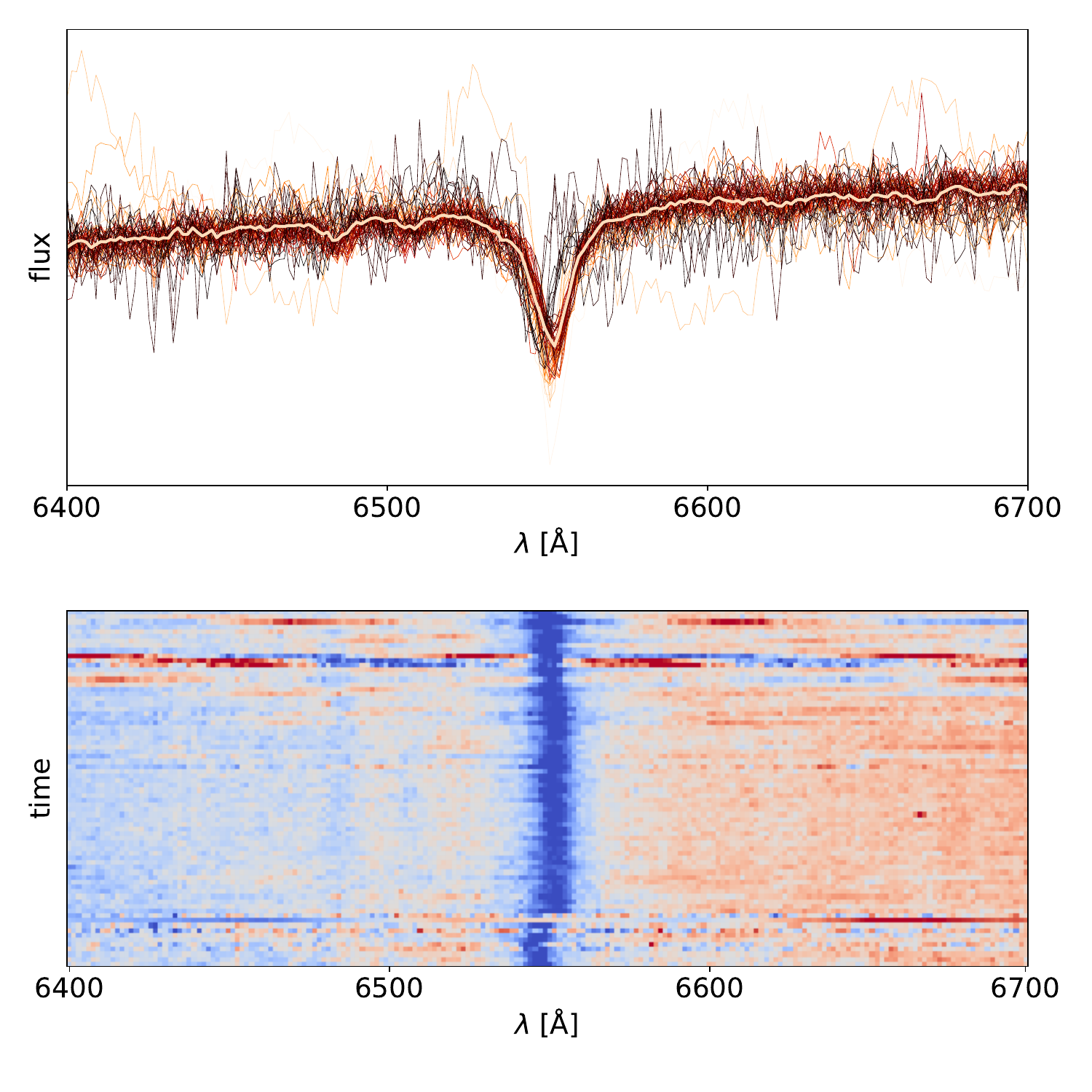}
\includegraphics[width=.30\textwidth]{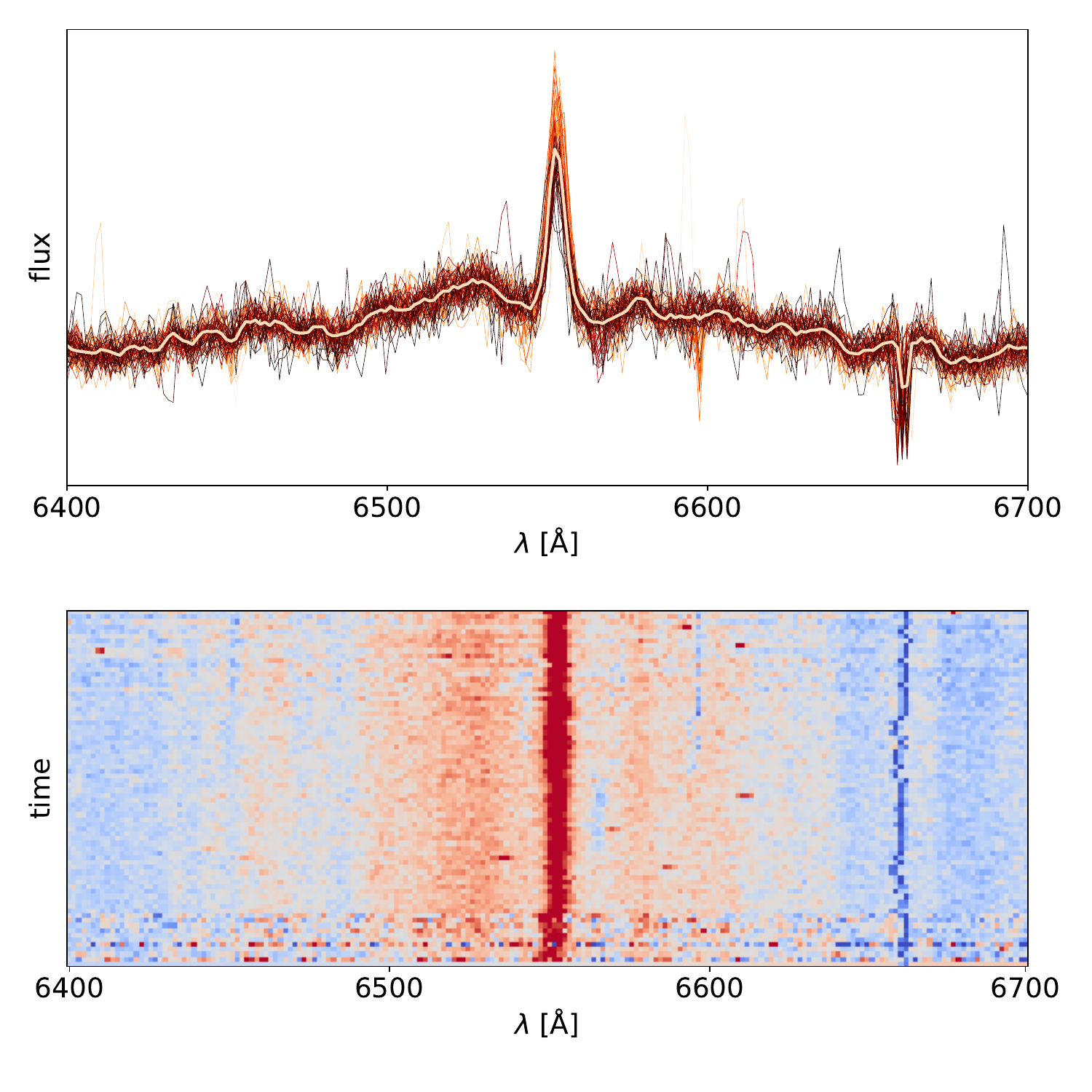}
\includegraphics[width=.30\textwidth]{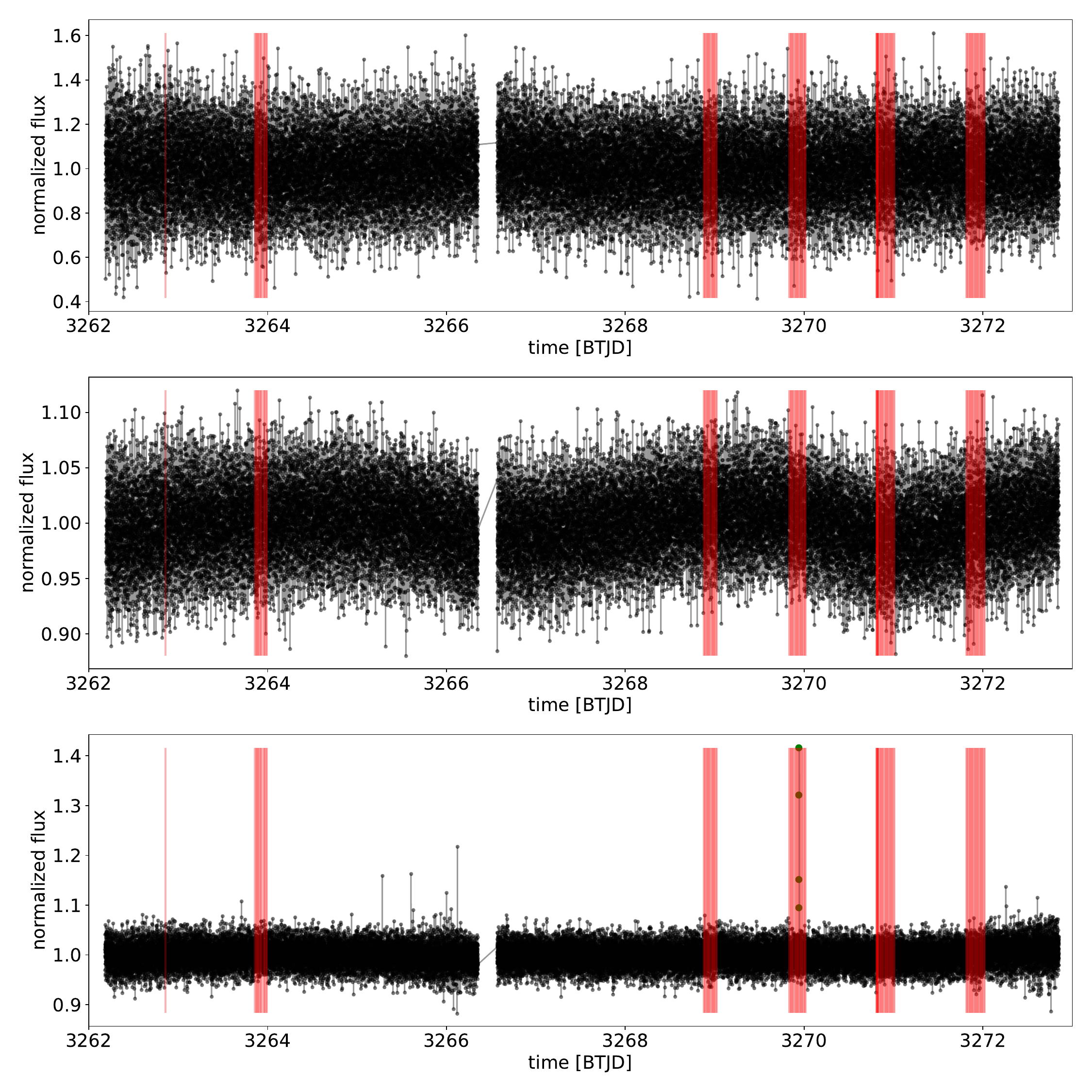}

    \caption{Example dynamic spectra and dynamic spectra from the Praesepe cluster showing the H$\alpha$ region. The continuous thick line indicates the averaged spectra. The last plot shows a sample of TESS light curves with the times of the spectroscopic observations marked with red boxes.}
    \label{fig:sample-praesepe}
\end{figure}

During the 10-day-long observing run in 2023 November, we used the OSMOS multi-object spectrograph on the MDM telescope (Arizona, US) 
with the moderate resolution VPH grism ($R\approx1600$). 
This setup yields a velocity resolution of $\approx180$km\,s$^{-1}$, enough to resolve larger coronal mass ejections  (typical CMEs have velocities ranging from a few hundred to a few thousand km\,s$^{-1}$) with good S/N while having decent temporal resolution. 
We had 12 hours of on-target time for the selected late-type stars in the Pleiades cluster 
and 15 hours of on-target time for the Praesepe cluster. During the MDM observing run the TESS also observed the cluster as part of the guest investigator program G06138. A sample of the resulting H$\alpha$ spectra is shown in Figures \ref{fig:sample-pleiades} and \ref{fig:sample-praesepe}.

We did not detect any obvious sign of coronal mass ejections or flares in the H$\alpha$ region. The total on-source time for the targets in the two clusters was close to 300 hours -- even if we discard the spectra with low signal-to-noise levels, based on the empirical estimations of \cite{2020MNRAS.494.3766O}, we would expect at least one CME detection. It is possible that the relatively low spectral resolution, ionization of the ejecta (with the ionization of the plasma it cannot be detected in H$\alpha$), and projection effects could have an influence on the results.
{In the future we will use the observations to improve the existing statistics and to refine the upper limits of stellar coronal mass ejections.}

\begin{acknowledgements}
This research was funded by the Hungarian National Research, Development and Innovation Office \'Elvonal grant KKP-143986. Authors acknowledge the financial support of the Austrian--Hungarian Action Foundation grants 112\"ou1. LK acknowledges the support of the Hungarian National Research, Development and Innovation Office grant PD-134784. KV and LK are supported by the Bolyai J\'anos Research Scholarship of the Hungarian Academy of Sciences. RMR is supported by the Heising-Simons Foundation's 51 Pegasi b Fellowship.

\end{acknowledgements}

\end{document}